\documentclass[prl,aps,twocolumn,showpacs]{revtex4}
\begin{document}
\title{Generation of NPT Entanglement from Nonclassical Photon Statistics}
\author{J. Solomon Ivan}
\email{solomon@imsc.res.in}
\author{N. Mukunda}
\email{nmukunda@cts.iisc.ernet.in}
\altaffiliation{Pemanent Address: Centre for High Energy Physics, Indian Institute of Science, Bangalore 560 012}
\author{R. Simon}
\email{simon@imsc.res.in} 
\affiliation{The Institute of Mathematical Sciences,  CIT Campus, Taramani, 
Chennai 600 113, India}

\date{April 03, 2006} 

\begin{abstract}
With a  product state of the form 
${\rho}_{\rm in} = {\rho}_{a} \otimes | 0 \rangle_b {_b} \langle 0 |$ 
as input,  the output two-mode state   $\rho_{\rm out}$ of the beam splitter is shown to be NPT 
whenever 
the photon number distribution (PND) statistics $\{\,p(n_a)\,\}$ associated with 
the possibly mixed state $\rho_a$ of the a-mode is 
antibunched or otherwise nonclassical, i.e., if $\{\,p(n_a)\,\}$ fails to respect any one of 
an infinite sequence of classicality conditions.
\end{abstract}
\maketitle

Considerable progress has been achieved in recent years towards an 
understanding of entanglement in the context of continuous variable bipartite 
systems\cite{expts, Simon, gauss, engen}. Phase space descriptions and 
Gaussian states have played a significant role in this progress.

While the quantum uncertainty principle places restrictions on the moments of 
the phase space variables, separability of a state places additional demands 
on these moments. Gaussian states are fully determined by their first and 
second moments, or variances, and so also is the issue of their separability.  
But non-Gaussian states which satisfy the separability demand on their 
variances could still be entangled by virtue of violation of one of the 
separability requirements on their higher order moments. It is only relatively 
recently that interest in this direction has started emerging\cite{moments}.

There is a useful connection between nonclassicality and inseparability,  
and the beam splitter plays an important role as a bridge 
between these two attributes. 
Asboth {\em et al.}\cite{asboth} have shown that  the output of a  beam splitter whose input 
is a 
 product state  
 ${\rho}_{\rm in} = {\rho}_{a} \otimes | 0 \rangle_b {_b} \langle 0 |$ 
 is  entangled if and only if the single-mode state $\rho_a$ 
 at the input is nonclassical.
 [Thus  
 any measure of entanglement of the output  state in this configuration 
is a computable  measure of nonclassicality of $\rho_a$, the entanglement potential 
(EP)\cite{asboth}]. For use in quantum information tasks, however, it is important to know if 
the entanglement 
generated in this manner is NPT or PPT.

 Two canonical manifestations of nonclassicality have been extensively studied in the 
quantum optics context: (1) squeezing\cite{squeeze}, and (2) antibunching  or 
sub-Poissonian fluctuation\cite{antibunch}, 
which is a particular manifestation of nonclassical photon statistics; these are  
 respectively nonclassicalities of the phase-sensitive and phase-insensitive types.  
Whereas the former  
has been well explored  as a source of entanglement in the context of Gaussian states, the 
same cannot be said 
in respect of the latter. 
 
The principal aim of this Letter is to show that with 
${\rho}_{\rm in} = {\rho}_{a} \otimes | 0 \rangle_b {_b} \langle 0 |$ 
 as input, the two-mode state after the beam splitter is definitely NPT 
if $\rho_a$ is nonclassical of the phase-insensitive type, that is if 
 the photon 
number distribution (PND) statistics $\{\,p(n_a)\,\}$ associated with  
 $\rho_a$ of the a-mode is antibunched, or possesses any other higher 
order nonclassicality. We prove this result in two stages: first for the restricted case of 
antibunched 
input $\rho_a$ wherein we exhibit a surprisingly simple witness; then 
in the case of arbitrary PND we take advantage of a complete characterization of 
nonclassicality based on the theory of moments\cite{nonclass,shohat-book}. 
 Our result and mode of proof extends  
 the domain of effectiveness of the partial transpose criterion for separability  
considerably beyond the 
traditional Gaussian regime.

Consider therefore a bipartite system, each part  being a  single mode 
radiation field, with
respective creation and annihilation operator pairs ${\hat{a}}^{\dagger}$, 
$\hat{a}$ and ${\hat{b}}^{\dagger}$, $\hat{b}$ acting on Hilbert spaces 
${\cal{H}}^a$, ${\cal{H}}^b$. 
Their only nonvanishing 
commutators are 
$[\hat{a},{\hat{a}}^{\dagger}] \, = \,[\hat{b},{\hat{b}}^{\dagger}]\,= \, 1$.
 The Fock or photon number states for the two modes provide convenient ONB's for 
${\cal{H}}^a$, ${\cal{H}}^b$ respectively: 
\begin{eqnarray}
|n_a \rangle & = & {(n_a!)}^{-1/2} {({\hat{a}}^{\dagger})}^{n_a} | 0 \rangle_{a}\,, \nonumber \\
|n_b \rangle & = & {(n_b!)}^{-1/2}{({\hat{b}}^{\dagger})}^{n_b} | 0 \rangle_{b}\,,~~ n_a  ,   
n_b ~ = ~ 0,1,2,\ldots\;\;\;
\end{eqnarray}
\noindent
The products $|n_a ,n_b \rangle$ $\equiv$ $|n_a\rangle \otimes 
|n_b \rangle$ form an ONB for ${\cal{H}}^{a} \otimes {\cal{H}}^{b}$.\
 The partial transpose of a bipartite state $\rho$ is 
the operator ${\tilde{\rho}}$ on ${\cal{H}}^a \otimes {\cal{H}}^b$ defined in this ONB 
 thus:
\begin{equation}
\langle n_a,n_b|{\tilde{\rho}}|{n_a}^{'},{n_b}^{'}\rangle 
         \,\equiv \, \langle n_a, n_b^{'}|\rho |n_a^{'} , n_b\rangle.
\end{equation}

\noindent
We have at our disposal this key fact\cite{test}:  if the partial transpose ${\tilde{\rho}}$ 
is not a valid density operator (NPT), i.e., if ${\tilde\rho} \not\geq 0$, 
then $\rho$ is definitely an entangled state. 

Clearly, a way to show that ${\rho}$ is NPT  
is to exhibit  an operator $\hat{A}$ of the bipartite system such 
that the `expectation value' of the {\em positive} operator 
$\hat{A^\dagger}\hat{A}$ in $\tilde{\rho}$
is negative.  We can then  conclude:
\begin{equation}
{\rm{Tr}} \,(\tilde{\rho}_{AB}\hat{A^\dagger}\hat{A}) < 0 , ~{\rm{for~ some}} ~ 
\hat{A} \, \Rightarrow \, \rho
~ {\rm{is~NPT}}.
\end{equation}
\noindent
Alternatively, and in a sense more directly, we may be able to find some 
principal submatrix of the matrix 
 $\langle n_a,n_b|{\tilde{\rho}}|{n_a}^{'},{n_b}^{'}\rangle$ 
representing $\tilde{\rho}$, i.e., a submatrix formed by 
intersections of any subset of {\em{rows of this matrix and the corresponding columns}},
such that this submatrix is not positive definite. Then again we can conclude:
\begin{eqnarray}
{\rm{A}} ~ {\rm{principal}} ~ {\rm{submatrix}} ~ {\rm{of}} ~
\tilde{\rho} \not\geq 0
 \, \Rightarrow \, \rho ~ {\rm{is ~NPT}}.\;\;\;
\end{eqnarray}
\noindent
We  use both strategies (3) and (4) in what follows.

\noindent
{\em Nonclassicality and Photon Statistics}: 
For a moment let us restrict attention to states $\rho_a$ of the $\hat{a}$-mode alone. 
The  separation of its (quantum) states into so called `classical' and 
`nonclassical' types is based on
the diagonal coherent state representation\cite{diag} of
${\rho}_a$:
\begin{eqnarray*}
{\rho}_a & = & \int_C {\pi}^{-1} {d^2 z_a}\, \phi(z_a) | z_a \rangle \langle z_a |\,, \nonumber \\
| z_a \rangle & = & e^{-{\frac{1}{2} {|z_a|}^2}} \sum_{n_a=1}^\infty
\frac{z_a^{n_a}}{\sqrt{n_a!}} |n_a \rangle\,,~~
\hat{a}| z_a \rangle  =  z_a | z_a \rangle\,.
\end{eqnarray*}
\noindent
The weight function $\phi(z_a)$ is in general a singular distribution of a
certain class. While it is real on account of hermiticity of ${\rho}_a$ and 
normalized to unity, it is in general not pointwise nonnegative over the complex
plane ${\cal C}$. Classical states are identified 
as follows.
\begin{equation}
{\rho}_{a} ~{\rm{`classical'}} ~ \Leftrightarrow ~
\phi(z_a) \geq 0 ~ {\rm{for}} ~ {\rm{all}} ~ z_a 
\in  {\cal C}.
\end{equation}
All other states are declared `nonclassical'.

All information about any state ${\rho}_{a}$ is captured 
by the expectation values of {\em all} possible hermitian observables  
$\hat{A}$
of the system, namely ${\rm{Tr}}\,({\rho}_{a} \hat{A} )$. 
 While any operator $\hat{A}$ can be written in the {\em normal ordered form} 
$\hat{A} = \sum_{j,k=0}^{\infty} c_{jk} \hat{a}^{\dagger j} \hat{a}^{k},\,$  
we limit ourselves 
to  observables that are `phase invariant' i.e., remain unchanged
under the map $\hat{a}$ $\rightarrow$ $e^{i\alpha} \hat{a}$,
${\hat{a}}^{\dagger}$ $\rightarrow$ $e^{-i\alpha} {\hat{a}}^{\dagger}$
for all $\alpha$. These are of the simple form
$\hat{A} = \sum_{j=0}^{\infty} c_{j} \hat{a}^{\dagger j} \hat{a}^{j}\,$,
 involving a single sum, and are just functions of the number operator $N_a=\hat{a}^{\dagger}
\hat{a}$. 

We can now ask for the way in which nonclassicality can manifest
itself if only measurements of such operators are carried out. Clearly 
these expectation values involve only the phase-averaged information contained 
in $\phi(z_a)$:
\begin{eqnarray*}
{\rm{Tr}} \,({\rho}_{a} \hat{A}) &  = &  {\int_{0}}^{\infty} d{I_a} P(I_a) 
\left( \sum_{j=0}^{\infty} c_{j} {I_a}^{j} \right), \nonumber \\
P(I_a) & = & {\int_{0}}^{2\pi} \frac{d \theta}{2 \pi} \phi\left({I_a}^{1/2} 
e^{i \theta}\right), ~~
 {\int_{0}}^{\infty} dI_a P(I_a)  =  1.
\end{eqnarray*}
\noindent 
While $P(I_a)$ is of course real, it may not be pointwise nonnegative. For a given
$P(I_a)$, the probabilities constituting the photon number distribution (PND)
are
\begin{eqnarray}
p(n_a)  \equiv  \langle n_a | {\rho}_{a} | n_a \rangle 
 =  {\int_{0}}^{\infty} dI_a P(I_a) \frac{e^{-I_a} {(I_a)}^{n_a}}{n_a !}\,.\;\;\;
\end{eqnarray}
\noindent
Whatever the nature of $P(I_a)$ may be, these probabilities are real 
nonnegative and add up to unity. 

We now present the following definition 
which will suffice for our purposes\cite{nonclass}:
\begin{eqnarray}
{\rho}_{a} \;\, {\rm{is}} \;\, {\rm{classical}} \;\, {\rm{in}} \;\,
{\rm{phase}} \; {\rm{invariant}} \;  {\rm{sense}} 
\;\,\Leftrightarrow\;\, P(I_a) \geq 0.\nonumber\\
\end{eqnarray}
\noindent
Otherwise ${\rho}_{a}$ is `nonclassical' in the phase invariant sense. This 
definition is a coarse-grained version of the earlier definition (5) 
given in terms of $\phi(z_a)$. Namely for the three mutually exclusive
situations
\begin{eqnarray}
&& (i) ~~ \phi(z_a) \geq 0 ~~ ({\rm{hence}} ~~
P(I_a) \geq 0)\,, \nonumber \\
&& (ii) ~~ \phi(z_a) \not\geq 0 ~~{\rm but}~~ P(I_a) \geq 0\,,
\nonumber \\
&& (iii) ~~ P(I_a) \not\geq 0 ~~ ({\rm{hence}} ~~ 
\phi(z_a) \not\geq  0)\,,
\end{eqnarray}
\noindent
while definition (5) would describe $(i)$ alone as `classical'
and $(ii)$, $(iii)$ as two levels of `nonclassicality', the phase-invariant
  definition (7) clubs $(i)$ and $(ii)$ together as 
`classical' and only $(iii)$ as `nonclassical'.

One very familiar and well known signature of phase invariant nonclassicality
is the antibunching condition 
\begin{eqnarray}
{\langle \Delta n_a \rangle}^2& - &\langle \ n_a \rangle  \equiv  
\langle \ {n_a}^2 \rangle - {\langle \ n_a \rangle}^2 - \langle \ n_a \rangle
\nonumber \\
& \equiv & {\int_{0}}^{\infty} dI_a P(I_a) {(I_a - \langle \ I_a \rangle)}^2
    \equiv  {( \Delta I_a )}^2 < 0,\;\;\;\; 
\end{eqnarray}
\noindent
which definitely implies $P(I_a) \not\geq 0$. 
But this is by no means the only such signature.
A complete set of necessary and 
sufficient conditions expressing the content of classicality, i.e., $P(I_a) \geq 0$, 
in terms of PND \{$p(n_a)$\} will be described later in this Letter..

\noindent
{\em Transpose Operation and Expectation Values for Operators}: While transpose operation on 
the density operator transcribes into `momentum reversal' in the Wigner phase space 
description\cite{Simon},  it is more convenient for our present purposes to transfer this 
operation 
directly onto operators and their expectation values.  
The transpose $\rho_a^T$ of the density operator $\rho_a$ (of the a-mode) is defined in the 
Fock basis through 
$\langle n|\rho_a^T | n '\rangle \equiv 
\langle n '| {\rho}_a| n\rangle$. Now the key point here is that in the Fock basis the 
operators $\hat{a}$,
${\hat{a}}^{\dagger}$ are represented by {\em real matrices}, so the transposition
operation for the matrix of ${\hat{a}}^{\dagger l}{\hat{a}}^{m}$ coincides 
with hermitian conjugation, and we have  
${\rm{Tr}}\,(\rho_a ^T{\hat{a}}^{\dagger j} {\hat{a}}^{k})
\,=\,{\rm{Tr}}\,({\rho}_a{\hat{a}}^{\dagger k} {\hat{a}}^{j})$.
 A linear combination  
$\sum_{j,\,k}{c_{j,k}}\hat{a}^{\dagger j}\hat{a}^{k}$, with complex coefficients 
$\{\,c_{j,k}\,\}$, would thus be taken under the transpose map  to 
$\sum_{j,\,k}{c_{j,k}}\hat{a}^{\dagger k}\hat{a}^{j}$, 
 in contradistinction to (the antilinear) hermitian conjugation which would have taken it to  
$\sum_{j,\,k}{c_{j,k}^{\ast}}\hat{a}^{\dagger k}\hat{a}^{j}$. It is the {\em linearity} of 
the transpose map  that permits its implementation on a subsystem, 
the b-mode, of a two-mode system through  
 \begin{eqnarray}
{\rm{Tr}}\,(\tilde{\rho}_{\rm out}{\hat{a}}^{\dagger j} {\hat{a}}^{k} 
{\hat{b}}^{\dagger l} {\hat{b}}^{m}) \,=\,
{\rm{Tr}}\,({\rho}_{\rm out}{\hat{a}}^{\dagger j} {\hat{a}}^{k} 
{\hat{b}}^{\dagger m} {\hat{b}}^{l}).
\end{eqnarray}
\noindent
This partial transpose relation  proves very useful.

\noindent
{\em Beam Splitter and Conversion of Nonclassicality into Entanglement}:  
The beam splitter is a passive, energy conserving system. Its unitary action 
$U$ on the 
two-mode Hilbert space ${\cal{H}}^{a} \otimes {\cal{H}}^{b}$
  can be viewed in two equivalent ways, and we will make use 
of both in succession. In the Heisenberg-type view, the effect of the beam splitter 
is to produce a $\pi/4$ rotation  in the $(\,\hat{a},\,\hat{b}\,)$ plane:  
\begin{eqnarray}
&& U \hat{a} {U}^{-1} = \frac{1}{\sqrt{2}} (\hat{a} + \hat{b})\,,
~~~ U \hat{b} {U}^{-1} = \frac{1}{\sqrt{2}} (\hat{b} - \hat{a})\,,
\nonumber \\
&& {U}^{-1} \hat{a} U = \frac{1}{\sqrt{2}} (\hat{a} - \hat{b})\,,
~~~ {U}^{-1} \hat{b} U = \frac{1}{\sqrt{2}} (\hat{b} + \hat{a})\,.
\end{eqnarray}
Alternatively, in the Schr\"{o}dinger-type view, wherein $U$ acts on the state 
rather than on dynamical variables,  the fact that  $U$ 
commutes with the operator 
$\hat{a}^{\dagger}\hat{a} + \hat{b}^{\dagger}\hat{b}$  and hence will not connect 
subspaces of the 
two-mode Hilbert space ${\cal{H}}^{a} \otimes {\cal{H}}^{b}$  differing in the total number 
of photons will prove very useful\cite{campos}. 

We will now employ the beam splitter in a simple 
situation, and later extend 
its use to a less restricted situation.

\noindent
{\em 1. The case of antibunched input}: 
Let us assume that our beam splitter is fed  with a product state 
\begin{equation}
{\rho}_{\rm in} = {\rho}_{a} \otimes | 0 \rangle_b {_b} \langle 0 |\,,
\end{equation}
with the b-mode in the vacuum state. After passing through the beam splitter we have the 
output state
 ${\rho}_{\rm out} = U {\rho}_{\rm in} {U}^{-1}\,$. 
To test the operator ${\tilde{\rho}}_{\rm out}$, 
the partial transpose of  ${\rho}_{\rm out}$, for positivity 
 we use the first strategy (3),  making the choice
$A = c_0 + c_1\hat{a}\hat{b}\,$. 

Arranging $c_1,\,c_2$ into a column $C$ so that $C^{\dagger}$ is the row 
vector $(c_1^{\ast},\,c_2^{\ast})$, we obtain  for the left hand side of (3) 
 \begin{equation}
{\rm{Tr}}\,({\tilde{\rho}}_{\rm out} A^{\dagger} A) = 
C^{\dagger}\left( \begin{array}{clcr}
      1  &  {\rm{Tr}}({\tilde{\rho}}_{\rm out} \hat{a} \hat{b} ) \\
 {\rm{Tr}}\,({\tilde{\rho}}_{\rm out} {\hat{a}}^{\dagger} {\hat{b}}^{\dagger} ) 
& {\rm{Tr}}\,({\tilde{\rho}}_{\rm out}{\hat{a}}^{\dagger}{\hat{a}}{\hat{b}}^{\dagger}
\hat{b})
\end{array} \right)C.
\end{equation}
 We now use the result (10) of the PT operation and eqs. (11), to relate these 
traces to corresponding expectation values in the original input state 
${\rho}_{a}$ of eq. (12):
\begin{eqnarray*}
{\rm{Tr}}\,({\tilde{\rho}}_{\rm out} \hat{a} \hat{b}) & = & {\rm{Tr}}\,({\rho}_{\rm out} \hat{a}
{\hat{b}}^{\dagger}) \nonumber \\
& = & {\rm{Tr}}\,({\rho}_{\rm in} {U}^{-1}{\hat{b}}^{\dagger} \hat{a} U)
\nonumber \\
& = & \frac{1}{2} {\rm{Tr}}\,({\rho}_{\rm in} ({\hat{b}}^{\dagger} + {\hat{a}}^{\dagger})
(\hat{a} - \hat{b})) \nonumber \\
& = & \frac{1}{2} {\rm{Tr}}\,({\rho}_{a} {\hat{a}}^{\dagger} \hat{a}) =
\frac{1}{2} \langle n_a \rangle\,.
\end{eqnarray*}
The other independent matrix element in (13) is, by (10) followed by
(11),
\begin{eqnarray*}
&&{\rm{Tr}}\,({\tilde{\rho}}_{\rm out}{\hat{a}}^{\dagger}  \hat{a}{\hat{b}}^{\dagger} 
\hat{b})  = 
{\rm{Tr}}\,({\rho}_{\rm out}{\hat{a}}^{\dagger} \hat{a} {\hat{b}}^{\dagger} \hat{b})
\nonumber \\
&&~~~~~~~~
=  {\rm{Tr}}\,({\rho}_{\rm in} {U}^{-1}{\hat{a}}^{\dagger} {\hat{b}}^{\dagger} \hat{a} \hat{b}U)
\nonumber \\
&&~~~~~~~~
= \frac{1}{4}{\rm{Tr}}\,({\rho}_{\rm in} ({\hat{a}}^{\dagger} - {\hat{b}}^{\dagger})
({\hat{a}}^{\dagger} + {\hat{b}}^{\dagger})
(\hat{a} - \hat{b})
(\hat{a} + \hat{b})) 
\nonumber \\
&&~~~~~~~~
=  \frac{1}{4}{\rm{Tr}}\,({\rho}_{a} {\hat{a}}^{\dagger} {\hat{a}}^{\dagger}\hat{a }
\hat{a}) =  \frac{1}{4} (\langle {n_a}^{2} \rangle - \langle n_a \rangle)\,.
\end{eqnarray*}
As a result, (13) reads:  
\begin{eqnarray*}
{\rm{Tr}}\,({\tilde{\rho}}_{\rm out} A^{\dagger} A) = 
C^{\dagger}\left( \begin{array}{clcr}
      1  &  \frac{1}{2}\langle n_a \rangle  \\
\frac{1}{2}\langle n_a \rangle &  \frac{1}{4} (\langle {n_a}^{2} 
\rangle - \langle n_a \rangle)
\end{array} \right)C.
\end{eqnarray*}
The determinant of this matrix is exactly the expression 
$\frac{1}{4} (\langle {\Delta n_a}^{2} \rangle - \langle n_a \rangle)$which 
appears in the antibunching condition (9). That is, if 
the input ${\rho}_{a}$ is antibunched, then 
${\rho}_{\rm out}$ is necessarily NPT. We have thus proved\\
{\em Theorem~1}: If a product state of the form   
${\rho}_{\rm in} = {\rho}_{a} \otimes | 0 \rangle_b {_b} \langle 0 |\,$, with antibunched  
$\rho_a$, is fed into a beam splitter, the output is definitely NPT entangled. That is, 
\begin{eqnarray}
(\langle {\Delta n_a}^{2} \rangle - \langle n_a \rangle) < 0 
~{\rm{for}} ~{\rho}_{a} &\Rightarrow& {\tilde{\rho}}_{\rm out} \not\geq 0 
\nonumber\\
&\Rightarrow& {\rho}_{\rm out} ~{\rm is} ~ {\rm{NPT}}.\;\;\;\;
\end{eqnarray}

\noindent
{\em 2. The case of  general nonclassical PND input}: 
In the previous case we allowed the input a-mode state ${\rho}_{a}$ to be any
antibunched state. Now we retain the product form (12) for the input two mode  
state, but choose for ${\rho}_{a}$ a more general phase invariant state determined completely
by some given PND probabilities \{$p(n_a)$\}:
\begin{equation}
{\rho}_{a}\left(\{p(n_a)\}\right) \equiv \sum_{n_a=0}^{\infty} p(n_a) | n_a \rangle \langle n_a |\,.
\end{equation}
The probabilities $p(n_a)$ are of course real nonnegative and normalized to
unity, but are otherwise free. Given them, the problem of inverting eq (6)
to obtain $P(I_a)$ is one of the classical moment problems, in fact the 
Stieltjes moment problem\cite{shohat-book} as we are concerned with the variable $I_a$ 
$\in$ $[0,\infty)$. In this case  $P(I_a)$ is uniquely
determined by \{$p(n_a)$\}; and the condition $P(I_a) \geq 0$ translates, in a {\em necessary 
and sufficient} manner,  into 
positivity conditions for a double hierarchy of real symmetric matrices
of increasing dimensions, formed out of the probabilities \{$p(n_a)$\}. These
are best described  in the following form\cite{nonclass}:
\begin{eqnarray}
q_{n_a} & = & n_a! p(n_a) \nonumber \\
{\rho}_{a}\left(\{p(n_a)\}\right) {\rm is} ~ &{\rm a}& ~ {\rm `classical'}
~{\rm  state}~ \nonumber\\
& \Leftrightarrow & ~ {\rm the}~ {\rm PND}~ \{p(n_a)\}~ {\rm is}
~{\rm `classical'}\,. \nonumber \\
& \Leftrightarrow & P(I_a) \geq 0  \nonumber \\
& \Leftrightarrow & L^{(N)},~~ {\tilde{L}}^{(N)} \geq 0 ~~ {\rm for} ~~
N=0,1,2, \ldots \nonumber
\end{eqnarray}
\begin{eqnarray}
&& L^{(N)} =
\left( \begin{array}{cccccc}
      {q_0} & {q_1} & {q_2} & {.} & {.} & {q_N}    \\
      {q_1} & {q_2} & {q_3} & {.} & {.} & {q_{N+1}} \\
      {.}   & {.}   & {.}   & {.} & {.}  & {.}      \\
      {.}   & {.}   & {.}   & {.} & {.}  & {.}      \\
      {q_N} & {q_{N+1}} & {q_{N+2}} & {.} & {.} & {q_{2N}} 
\end{array} \right)\,,
\nonumber \\
&& {\tilde{L}}^{(N)} =
\left( \begin{array}{cccccc}
      q_1 & q_2 & q_3 & . & . & q_{N+1}    \\
      q_2 & q_3 & q_4 & . & . & q_{N+2} \\
      .   & .   & .   & . & .  & .      \\
      .   & .   & .   & . & .  & .      \\
      q_{N+1} & q_{N+2} & q_{N+3} & . & . & q_{2N+1} 
\end{array} \right).
\end{eqnarray}
Now pass the two-mode product state 
\begin{equation}
{\rho}_{\rm in} = {\rho}_{a}(\{p(n_a)\}) \otimes | 0 \rangle_b {_b}\langle 0 |\,,
\end{equation}
through the beam splitter. The output  state is
\begin{eqnarray*}
&&{\rho}_{\rm out}  =  U {\rho}_{\rm in} {U}^{-1} \nonumber \\
&&~~
 =  U \sum_{n_a=0}^{\infty} \frac{p(n_a)}{n_a!} {({\hat{a}}^{\dagger})}^{n_a}
|0,0 \rangle \langle 0,0 | {(\hat{a})}^{n_a} {U}^{-1}  
\nonumber \\
&&~~
 =  \sum_{n_a=0}^{\infty} \frac{p(n_a)}{{2}^{n_a} n_a!} 
{({\hat{a}}^{\dagger} + {\hat{b}}^{\dagger})}^{n_a}
| 0 ,0 \rangle \langle 0 , 0 | {(\hat{a} +\hat{b})}^{n_a} 
\nonumber \\ 
&&~~
 =  \sum_{n_a=0}^{\infty}  \frac{p(n_a)n_a!}{{2}^{n_a}} 
\sum_{r,s=0}^{n_a} 
\frac{|r,{n_a}-r\rangle \langle s,{n_a}-s|}{\sqrt{r!({{n_a}-r})!s!({{n_a}-s})!}}\,.
\end{eqnarray*}
The general matrix element of this density matrix is
\begin{eqnarray*}
&&\langle {n_a}^{'}, {n_b}^{'} | {\rho}_{\rm out} | n_a, n_b 
\rangle \nonumber\\
&&~~~~~
= \delta_{{n_a}^{'} +{n_b}^{'}, n_a + n_b}
\frac{(n_a + n_b)! p(n_a + n_b)}
{2^{n_a + n_b}\sqrt{{n_a}^{'}!{n_b}^{'}!n_a!n_b!}}\,.
\end{eqnarray*}
The matrix elements of the partial transpose ${\tilde{\rho}}_{\rm out}$ are 
obtained simply by 
interchanging $n_b$ and ${n_b}^{'}$:
\begin{eqnarray}
&&\langle {n_a}^{'}, {n_b}^{'} | {\tilde{\rho}}_{\rm out} | n_a, n_b 
\rangle \nonumber\\
&&~~~~~
= \delta_{{n_a}^{'} +n_b,n_a + {n_b}^{'}}
\frac{{q}_{n_a + {n_b}^{'}}}
{2^{n_a + {n_b}^{'}}\sqrt{{n_a}^{'}!{n_b}^{'}!n_a!n_b!}}\,.
\end{eqnarray}

Now we test for possible lack of positivity of $\tilde{\rho}_{{\rm out}}$ by using the second 
strategy (4).
Consider first the principal submatrix of (18) obtained by setting 
${n_b}^{'} = {n_a}^{'}$, $n_b = n_a\,$.
 This submatrix $H^{(N)}$ has elements
\begin{displaymath}
H^{(N)}_{{n_a}^{'},n_a} \equiv
\langle {n_a}^{'}, {n_a}^{'} | {\tilde{\rho}}_{{\rm out}} | n_a, n_a \rangle
= \frac{{q}_{n_a + {n_a}^{'}}}{2^{n_a + {n_a}^{'}} {n_a}^{'}!n_a!}.
\end{displaymath}
Clearly, this  (N+1) dimensional submatrix coincides with $L^{(N)}$ modulo 
congruence by a diagonal matrix $D^{(N)}$: 
\begin{eqnarray}
H^{(N)} &=& D^{(N)} L^{(N)}  D^{(N)},\nonumber\\
D^{(N)}_{n_a,\,{n_a}^{'}} &=& 
(\,2^{{n_a}}{n_a}!\,)^{-1}\,  \delta_{n_a,\,{n_a}^{'}}\,.
\end{eqnarray}
Consider similarly the principal submatrix of (18) corresponding to the row 
(and column) choices ${n_b}^{'} ={n_a}^{'} +1$ and $n_b = n_a +1$. This 
submatrix $\tilde{H}^{(N)}$ has elements
\begin{eqnarray*}
\tilde{H}^{(N)}_{{n_a}^{'},n_a} &\equiv&
\langle {n_a}^{'}, {n_a}^{'}+1 | {\tilde{\rho}}_{out} | n_a, n_a + 1 \rangle
\nonumber\\
&=& \frac{{q}_{n_a + {n_a}^{'}+1}}{2^{n_a + {n_a}^{'}+1} \sqrt{({n_a}^{'}+1)!
{n_a}^{'}! (n_a+1)!n_a!}}.
\end{eqnarray*}
This  (N+1) dimensional submatrix is seen to coincide with $\tilde{L}^{(N)}$ 
modulo congruence by a diagonal matrix: 
\begin{eqnarray}
\tilde{H}^{(N)} &=& \tilde{D}^{(N)} \tilde{L}^{(N)}\tilde{D}^{(N)}\nonumber\\
\tilde{D}^{(N)}_{n_a,\,{n_a}^{'}} &=& 
\left(\,2^{{n_a}+ \frac{1}{2}}\sqrt{({n_a}+1)!{n_a}!}\,\right)^{-1}
    {\delta_{n_a,\,{n_a}^{'}}}.\;
\end{eqnarray}

We see that in both (19) and (20), the exhibited  principal submatrices of 
${\tilde{\rho}}_{{\rm out}}$ are essentially the matrices 
$L^{(N)}$  and ${\tilde{L}}^{(N)}$ 
(apart from trivial congruences) appearing in the result
(16) of the moment problem to ensure positivity of $P(I_a)$\cite{nonclass}. 
 Thus if $\rho_a(\{p(n_a)\}$ is nonclassical, i.e., if $P(I_a) \not\ge 0$, 
  then one of the matrices $L^{(N)}$, ${\tilde{L}}^{(N)}$ will necessarily fail to be 
positive semidefinite,  for some $N$, as seen from (16), rendering one of the above 
submatrices of ${\tilde{\rho}}_{{\rm out}}$ nonpositive. Thus, we arrive at 

\noindent
{\em Theorem 2}: If the input PND statistics  $\{p(n_a)\}$ has {\em any}  (phase-invariant) 
nonclassicality, 
then the output two-mode state emerging after  the beam splitter is NPT:
\begin{eqnarray}
L^{(N)} ~&{\rm or}& ~{\tilde{L}}^{(N)} \not\geq  0
 ~ {\rm for} ~ {\rm some} ~ N \nonumber\\
& \Leftrightarrow & 
{\rho}_{a}\left(\{p(n_a)\}\right) 
{\rm {correspondingly}} ~
{\rm {`nonclassical'}}  \nonumber\\
 & \Rightarrow & {\rm {corresponding}}
~ {\rm {submatrices}} ~ {\rm of}~
{\tilde{\rho}}_{\rm out} \not\geq 0 \nonumber \\
~ & \Rightarrow & ~ {\rho}_{\rm out}
~ {\rm is ~NPT ~entangled}.
\end{eqnarray}

To conclude, the output nonclassicality, the transform by the beam splitter  
of the  nonclassicality residing locally
in the input a-mode photon statistics, manifests itself nonlocally as NPT 
entanglement.  
It is rather  remarkable that the infinite set of matrices in (19), (20) 
whose positivity is equivalent to the  positivity of $\tilde{\rho}_{\rm out}$ 
should essentially coincide  with the infinite set of matrices in (16) designed,  
 with the help of the classical Stieltjes moment problem\cite{shohat-book}, to capture the 
classicality condition $P(I_a)\ge0$. 


\end{document}